\documentclass[aps,prb,twocolumn,showpacs,preprintnumbers,amsmath,amssymb,superscriptaddress]{revtex4}
\usepackage{graphicx}

\begin{document}

\title{Evidence for time-reversal symmetry breaking in superconducting
PrPt$_4$Ge$_{12}$}

\author{A.\ Maisuradze}
\email{alexander.maisuradze@psi.ch}
\affiliation{Laboratory for Muon Spin Spectroscopy, Paul Scherrer Institute, CH-5232 Villigen PSI, Switzerland}
\author{W.\ Schnelle}
\affiliation{Max Planck Institute for Chemical Physics of Solids, N\"othnitzer Str.\ 40, 01187 Dresden, Germany}
\author{R.\ Khasanov}
\affiliation{Laboratory for Muon Spin Spectroscopy, Paul Scherrer Institute, CH-5232 Villigen PSI, Switzerland}
\author{R.\ Gumeniuk}
\affiliation{Max Planck Institute for Chemical Physics of Solids, N\"othnitzer Str.\ 40, 01187 Dresden, Germany}
\author{M.\ Nicklas}
\affiliation{Max Planck Institute for Chemical Physics of Solids, N\"othnitzer Str.\ 40, 01187 Dresden, Germany}
\author{H.\ Rosner}
\affiliation{Max Planck Institute for Chemical Physics of Solids, N\"othnitzer Str.\ 40, 01187 Dresden, Germany}
\author{A.\ Leithe-Jasper}
\affiliation{Max Planck Institute for Chemical Physics of Solids, N\"othnitzer Str.\ 40, 01187 Dresden, Germany}
\author{Yu.\ Grin}
\affiliation{Max Planck Institute for Chemical Physics of Solids, N\"othnitzer Str.\ 40, 01187 Dresden, Germany}
\author{A.\ Amato}
\affiliation{Laboratory for Muon Spin Spectroscopy, Paul Scherrer Institute, CH-5232 Villigen PSI, Switzerland}
\author{P.\ Thalmeier}
\affiliation{Max Planck Institute for Chemical Physics of Solids, N\"othnitzer Str.\ 40, 01187 Dresden, Germany}

\begin{abstract}
Zero and longitudinal field muon spin rotation ($\mu$SR) experiments
were performed on the superconductors PrPt$_4$Ge$_{12}$ and
LaPt$_4$Ge$_{12}$. In PrPt$_4$Ge$_{12}$ below $T_c$ a spontaneous
magnetization with a temperature variation resembling that of the
superfluid density appears. This observation implies time-reversal
symmetry (TRS) breaking in PrPt$_4$Ge$_{12}$ below $T_c = 7.9$\,K. This
remarkably high $T_c$ for an anomalous superconductor and the weak and
gradual change of $T_c$ and of the related specific heat anomaly upon La
substitution in La$_{1-x}$Pr$_x$Pt$_4$Ge$_{12}$ suggests that the TRS
breaking is due to orbital degrees of freedom of the Cooper pairs.
\end{abstract}

\pacs{76.75.+i, 74.70.Dd, 74.25.Ha}

\maketitle

\section{Introduction}

The large family of filled skutterudite compounds $RT_4X_{12}$ ($R$ =
rare-earth, actinides, alkaline-earth, and alkali metals; $T$ = Fe, Ru,
Os; $X$ = P, As, Sb) displays an astonishing diversity of physical
properties among which superconductivity represents a particularly
complex one. Within more than 20 isostructural skutterudites known up to
now a perplexing multitude of conventional and unconventional
superconducting phases has been
observed.\cite{Meisner8185,Uchiumi99,EDBauer02,Shirotani03,Shirotani05}
In large part, the filler cations $R$ which are embedded in the
polyanionic [$T_4X_{12}$] host structure have a significant influence on
these properties. Superconducting members of this family containing Pr
have attracted considerable interest with PrOs$_4$Sb$_{12}$ being the
most prominent one. While LaOs$_4$Sb$_{12}$ (critical temperature $T_c$
= 0.74\,K) is found to obey the classical BCS-theory, PrOs$_4$Sb$_{12}$
($T_c$ = 1.85\,K) exhibits heavy-fermion behavior and unconventional
superconductivity\cite{Sugawara02,Parker08,Maple07both} with
time-reversal symmetry (TRS) breaking.\cite{Aoki03,Chia04} Moreover,
multiple superconducting phases and order parameters with nodes have
been detected.\cite{Maple07both} Recent research efforts show that these
phenomena depend on a subtle interplay of the crystal electric field
(CEF) acting on the Pr$^{3+}$ ion together with the hybridization of the
$f$-shell with the conduction electrons of the host. This is also found,
e.g., for the $R$[Fe$_4$P$_{12}$] system, where LaFe$_4$P$_{12}$ and
YFe$_4$P$_{12}$ are conventional superconductors and isovalent
substitution by Pr leads to an antiferro-quadrupolar ground state with
heavy electron masses.\cite{Sugawara02,Nakai05,Magishi05}

Recently, we investigated the properties of a different family of
compounds with a filled skutterudite structure based on platinum and
germanium, $R$Pt$_4$Ge$_{12}$ ($R$ = Sr, Ba, La, Ce, Pr, Nd,
Eu).\cite{Gumeniuk08a} The compounds with Sr and
Ba,\cite{Gumeniuk08a,Bauer07a} Th,\cite{Kaczorowski08} and with La and
Pr\cite{Gumeniuk08a,Maisuradze09b} are superconductors. These latter
compounds have the highest $T_c$ among the [Pr$_4$Ge$_{12}$]
skutterudites of 8.3\,K and 7.8\,K, respectively. In addition to the
surprisingly high $T_c$ of PrPt$_4$Ge$_{12}$ its superconducting energy
gap has point nodes, as has been demonstrated by specific heat as well
as muon spin rotation measurements down to very low reduced temperatures
($T/T_c \leq 0.005$).\cite{Maisuradze09b}

An analysis of the temperature variation of the superfluid density has
shown that the data can be well described by three selected gap
functions, of which two are compatible with the thermodynamic
data.\cite{Maisuradze09b} One of the remaining functions, $|\Delta| =
\Delta_0|\hat{k}_x - i\hat{k}_y|$, has been favored to describe the
\textit{unconventional} superconducting low-field (B) phase of
PrOs$_4$Sb$_{12}$, for which TRS
breaking\cite{Maple07both,Aoki03,Chia0305combi,Chia04} is observed and
has been discussed in connection with spin-triplet
pairing.\cite{Higemoto07} Moreover, the gap-to-$T_c$ ratios
$\Delta_0/k_BT_c$ of the two Pr superconductors are similar. While these
aspects of the superconducting states of PrPt$_4$Ge$_{12}$ and
PrOs$_4$Sb$_{12}$ are similar, the CEF splitting distinguishes the
compounds. Having the same nonmagnetic singlet ground state $\Gamma_1$,
in PrOs$_4$Sb$_{12}$ the first excited triplet $\Gamma_4^{(2)}$ ($E/k_B
\simeq 7-10$\,K)\cite{Maple07both} strongly hybridizes with the ground
state and the conduction electrons, generating the heavy-fermion state.
In PrPt$_4$Ge$_{12}$ the first excited CEF state is a different triplet
($\Gamma_4^{(1)}$ in $T_h$ notation). The $\Gamma_1$--$\Gamma_4^{(1)}$
splitting is huge
(120--130\,K),\cite{Gumeniuk08a,Toda08a,Goremychkin09abs} allowing for a
$T_c$ only little less than for LaPt$_4$Ge$_{12}$. No heavy-electron
states are present at the Fermi surface of PrPt$_4$Ge$_{12}$, as can be
concluded from thermodynamic data.\cite{Gumeniuk08a}

TRS breaking can lead to the appearance of a small magnetic moment
of the superconducting condensate due to spin or orbital degrees of
freedom of the Cooper pairs.\cite{Sigrist91} Muon spin rotation
($\mu$SR) successfully detected this field in a number of
\textit{unconventional} and \textit{spin-triplet}
superconductors.\cite{Heffner90,Luke98,Aoki03,Higemoto07,Hillier09}
Here, we report on detailed zero magnetic field (ZF) $\mu$SR
experiments in PrPt$_4$Ge$_{12}$ and LaPt$_4$Ge$_{12}$. The absolute
value and the mechanism of ZF muon depolarization above $T_c$ in
PrPt$_4$Ge$_{12}$ are similar to that reported for
PrOs$_4$Sb$_{12}$. Below $T_c$ = 7.8\,K a spontaneous magnetization
resembling the temperature dependence of the superfluid density was
observed for PrPt$_4$Ge$_{12}$. No such anomaly is detected for
LaPt$_4$Ge$_{12}$. The magnitude of this magnetization is of the
same order as that reported for PrOs$_4$Sb$_{12}$ and other
superconductors with TRS
breaking.\cite{Luke98,Heffner90,Aoki03} Due to the
contrasting behaviors of the La and Pr compound and in order to
elucidate the origin of the TRS breaking we synthesized samples of
the solid solution La$_{1-x}$Pr$_x$Pt$_4$Ge$_{12}$ and studied the
variation of $T_c$ and of the specific heat anomaly.

\section{Experimental}

The preparation procedures of the La$_{1-x}$Pr$_x$Pt$_4$Ge$_{12}$
samples are similar to that described previously.\cite{Gumeniuk08a} The
end member samples have residual resistance ratios
$\rho_\mathrm{300\,K}$/$\rho_0 \geq 30$ and the PrPt$_4$Ge$_{12}$ sample
showed crystallites up to 2\,mm size. The zero and longitudinal field
(ZF \& LF) $\mu$SR experiments were performed on the DOLLY spectrometer
at the $\pi$E1 beam line at the Paul Scherrer Institute (Villigen,
Switzerland). In addition, a powdered sample of PrPt$_4$Ge$_{12}$ was
measured in ZF on the GPS spectrometer at the $\pi$M3 beam line. The
samples were cooled in ZF or LF down to 1.5\,K and $\mu$SR spectra were
taken as a function of temperature. During ZF measurements, an active
magnetic-field compensation with three orthogonal couples of Helmholtz
coils was used in order to reduce the field at the sample to values
lower than $3 \times 10^{-6}$\,T. Typical counting statistics were $12
\times 10^6$ positron events per each particular data point.
Magnetization was measured in a commercial SQUID magnetometer.

\section{Results and discussion}

\begin{figure}[tb]
\includegraphics[width=0.99\linewidth]{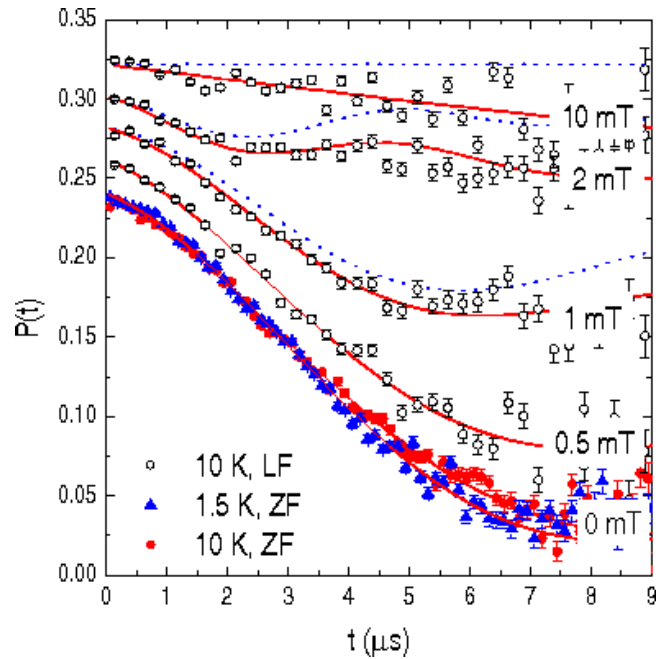}
\caption{(Color online)
Zero field $\mu$SR time spectra at 1.5\,K
($\blacktriangle$) and 10\,K ($\bullet$) for PrPt$_4$Ge$_{12}$. The
corresponding solid red lines are fits to the data according
to Eq.\,(\ref{eq:TimeSpectr}).
($\circ$) Spectra measured with longitudinal fields (LF) of
0.5\,mT, 1\,mT, 2\,mT and 10\,mT at 10\,K and corresponding fit with
Eq.\,\ref{eq:TimeSpectr} (the solid red lines). The blue dashed lines
are simulation of the spectra assuming
only the static field distribution (i.e.\ $\lambda_d = 0$).
For better visualization each LF spectrum is shifted by 0.02 units.
\label{fig:spectra}}
\end{figure}

Figure \ref{fig:spectra} shows ZF and LF $\mu$SR time spectra for
PrPt$_4$Ge$_{12}$ at 1.5\,K and 10\,K.
The spectra of several histograms were fitted
simultaneously. Each histogram is described by the function:
\begin{equation}\label{eq:HistSpectr}
N(t) = N_0 \exp(-t/\tau) (1+AP(t)) + B,
\end{equation}
where $\tau$ = 2.197019\,$\mu$s is the muon lifetime, $N_0$ a
proportionality coefficient, $B$ the background, $A$ the asymmetry, and
$P(t)$ the muon depolarization function. Preliminary fits showed that
the ZF muon depolarization is well described by a Kubo-Toyabe
depolarization function reflecting a static Voigt field distribution,
i.e.\ describing two mechanisms for the profile, one producing a Gaussian
distribution and one producing a Lorentzian distribution.

The zero and longitudinal field $\mu$SR depolarization functions for the
static Voigt field distribution were calculated using the general
formula derived by R.\ Kubo, (Eq.\,21 of Ref.\ \onlinecite{Kubo81}).
For Voigt-like functions this equation can be reformulated as follows:
\begin{equation}\label{eq:KuboGeneral}
P_G(t) = 1-\frac{2Q'(t)}{\omega_0^2t}[\cos{\omega_0t}-j_0(\omega_0t)]-
2\int_0^t\frac{Q'(s)}{s}\frac{j'_0(\omega_0s)}{\omega_0}ds.
\end{equation}
Here, $j_0(x)=\sin(x)/x$, $\omega_0 = \gamma_\mu B$ is the Larmor
frequency corresponding to the applied longitudinal
field $B$. For the Voigt function
$Q(t)=\exp(-\frac{1}{2}\sigma^2t^2-\lambda t)$,
where $\sigma^2/\gamma_{\mu}^2$
is the second moment of the Gaussian distribution and
$\lambda/\gamma_{\mu}$ is the half-width at half-maximum of the Lorentzian
distribution, and finally the prime denotes the derivative.\cite{Kubo81}
In the limit of $\omega_0 \rightarrow 0$ (i.e.\ in the ZF situation),
Eq.\,\ref{eq:KuboGeneral} converges
to the ``golden formula'' of Kubo:\cite{Kubo97}
\begin{equation}\label{eq:ZFKuboGeneral}
P_{G,ZF}(t) = \frac{1}{3} + \frac{2}{3}[Q(t)+Q'(t)t] ,
\end{equation}
and for the case of $Q(t)=\exp(-\frac{1}{2}\sigma^2t^2-\lambda t)$
one finally gets the equation:
\begin{equation}\label{eq:VoigtKuboToyabe}
P_{G,ZF,V}(t) = \frac{1}{3}+\frac{2}{3}(1-\sigma^2t^2-\lambda t)
\exp\left(-\frac{1}{2}\sigma^2t^2-\lambda t\right).
\end{equation}

Actually, a closer look at the data indicates that the ZF and
LF data are best fitted using the depolarization function:
\begin{equation} \label{eq:TimeSpectr}
P(t) = P_G(t)\exp(-\lambda_d t),
\end{equation}
where $\lambda_d \simeq 0.020$\,$\mu$s$^{-1}$ (which is practically
temperature independent) is a dynamical muon depolarization which does
not decouple up to fields of 20\,mT. Such dynamical depolarization is
best seen in LF experiments. In Fig.\,\ref{fig:spectra}, the fits
obtained with Eq.\,\ref{eq:TimeSpectr} are shown with the solid red curves.
The best fit to the data is obtained
with field independent parameters $\sigma$ = 0.173\,$\mu$s$^{-1}$,
$\lambda$ = 0.029\,$\mu$s$^{-1}$, and $\lambda_d$ = 0.020\,$\mu$s$^{-1}$.
The total asymmetry $A = 0.242$ was fixed during the fitting procedure.
The small dynamic contribution $\lambda_d$ to the relaxation is obvious
only when comparing the LF spectra with the depolarization curves
calculated for the case of a static only field distribution (see the
blue dotted curves in Fig.\,\ref{fig:spectra}). Note that for such small
values of $\lambda$ and $\lambda_d$, as in the present case, these
parameters are strongly correlated in the ZF spectra. However, LF
experiments allow us to disentangle this correlation.
Zero and longitudinal field experiments suggest the presence of static
muon depolarization predominantly from nuclear moments of the Pr, Pt, or
Ge isotopes. Note that usually for a depolarization due to nuclear
moments, one assumes a Gaussian field-distribution (i.e.\ a situation
with $\lambda = 0$). However, a pure Gaussian field distribution is an
approximation and does not take into account, for example, the presence
of different isotopes with different nuclear moments (as for Pt and Ge).
It is therefore likely that the field distribution due to nuclear
moments is not purely Gaussian in our case. Note also that the main
conclusions concerning the temperature dependence of the muon
depolarization (see below) do not depend on the exact static field
distribution assumed. Decoupling of this static field is well described
with the general expression given in Eq.\,\ref{eq:TimeSpectr}.\cite{Kubo81}
The small dynamic contribution $\lambda_d$ which does not decouple up to
fields of 20\,mT is presumably due to some additional spin-lattice
relaxation mechanisms.

\begin{figure}[!t]
\includegraphics[width=0.99\linewidth]{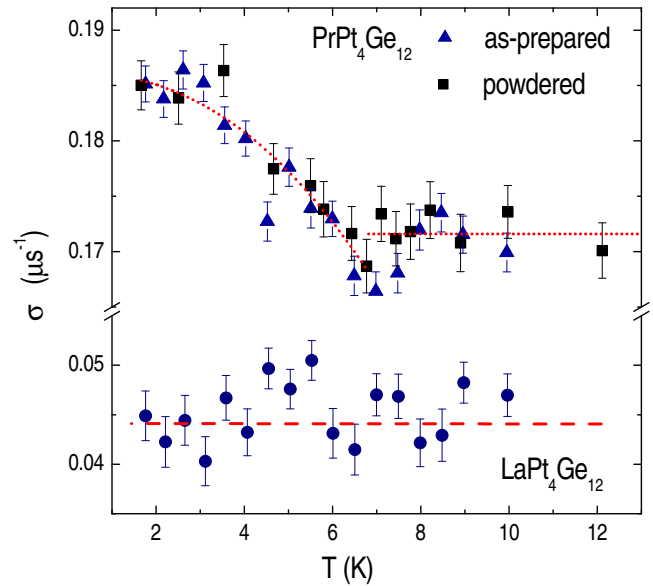}
\caption{(Color online) Temperature dependence of the muon
depolarization rate $\sigma$ in ``as-prepared'' ($\blacktriangle$)
and powdered ($\blacksquare$) samples of PrPt$_4$Ge$_{12}$
and in ``as-prepared'' LaPt$_4$Ge$_{12}$ ($\bullet$) as obtained
by Eq.\,(\ref{eq:TimeSpectr}). The lines are guides to the eye
(see text).\label{fig:RlxVsT}}
\end{figure}

When studying the temperature dependence of the parameters $\sigma$ and
$\lambda$, we first noticed that $\lambda$ is practically independent of
temperature.
In a second step, $\lambda$ was fixed to its average value
0.029\,$\mu$s$^{-1}$ and solely $\sigma$ was kept free. Figure
\ref{fig:RlxVsT} presents $\sigma(T)$ recorded in ZF for
PrPt$_4$Ge$_{12}$ and LaPt$_4$Ge$_{12}$, measured on two different
spectrometers.
For the Pr compound, above $T_c$ = 7.8\,K $\sigma$ is independent of
temperature, as expected for depolarization due to nuclear moments.

However below $T_c$ one can observe a clear increase of
the muon relaxation rate with decreasing temperature. Just below
$T_c$ the data systematically decrease below the normal-state level
showing a small dip. The rise of $\sigma$ starts only below $\approx
6.5$\,K. There is no indication for a phase transition at this
temperature from other measurements as, e.g., specific heat or the
superfluid density.\cite{Gumeniuk08a,Maisuradze09b}

The corresponding ZF depolarization rates $\sigma(T)$ for
LaPt$_4$Ge$_{12}$ with $T_c = 8.3$\,K (obtained via
Eq.\,\ref{eq:TimeSpectr} with $\lambda = \lambda_d = 0$ and free parameter
$\sigma$) are small and temperature-independent (see Fig.\
\ref{fig:RlxVsT}).\cite{RemarkFit} No anomaly is resolved at $T_c =
8.3$\,K. $\sigma$ for LaPt$_4$Ge$_{12}$ is substantially smaller than in
PrPt$_4$Ge$_{12}$, indicating that in PrPt$_4$Ge$_{12}$ the dominant part
of the relaxation is due to the presence of $^{141}$Pr nuclei.

A similarly strong (nearly the same value of $\sigma$)
hyperfine-enhanced nuclear muon depolarization was observed in the
isostructural PrOs$_{4-x}$Ru$_x$Sb$_{12}$ compounds.\cite{LeiShu07}
The authors explain the relaxation by a Van-Vleck-like admixture of
magnetic excited CEF states into the nonmagnetic $\Gamma_1$ ground
state of Pr$^{3+}$ by the nuclear hyperfine coupling. This
hybridization strongly increases the strength of the interactions
between the $^{141}$Pr nuclear spins and the muon spins as well as
within the $^{141}$Pr nuclear spin system.\cite{Bleaney73} Lei Shu
\textit{et al.} observe that this relaxation is \textit{dynamic} due
to the relatively low energy ($E/k_B \simeq 7$--10\,K for $x$ = 0) of
the first exited CEF level $\Gamma_4^{(2)}$ of Pr$^{3+}$ with a
spin-spin correlation time $\tau_c \simeq 0.2$--0.6 $\mu$s$^{-1}$.
For PrPt$_4$Ge$_{12}$, the first exited CEF level $\Gamma_4^{(1)}$ is
found at $E/k_B \simeq
120$--130\,K,\cite{Gumeniuk08a,Toda08a,Goremychkin09abs} in agreement
with our observation of a quasi-static nuclear magnetism of Pr.
The very small population of all exited CEF states at $T \simeq T_c$ =
7.8\,K is the reason for this behavior and the negligible Cooper-pair
breaking in PrPt$_4$Ge$_{12}$. The origin of the additional dynamic
relaxation $\lambda_d$ in the present case is unknown.
To reduce the magnitude of $\lambda_d$ in LF the LF Larmor precession
frequency should exceed the characteristic fluctuations of magnetic
field probed in the sample.\cite{Uemura85}
Since it does not decouple up to 20\,mT one estimates the
characteristic fluctuation frequency larger than $\nu> 2\pi
\gamma_\mu\cdot 0.02$ = 17\,MHz.

\begin{figure}[tb]
\includegraphics[height=0.99\linewidth,angle=90]{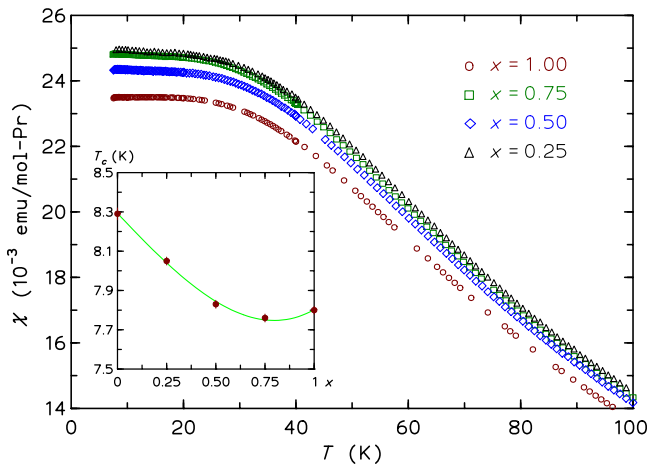}
\caption{(Color online) Magnetic susceptibility of
Pr$_x$\-La$_{1-x}$á-Pt$_4$Ge$_{12}$ samples in $\mu_0H$ = 0.1\,T. The
inset shows the superconducting $T_c$ ($\mu_0H$ = 2\,mT) \textit{vs}
the nominal Pr-content $x$.\label{fig:mag}}
\end{figure}

Figure \ref{fig:mag} shows the susceptibility measurements for the
series of La$_{1-x}$Pr$_x$Pt$_4$Ge$_{12}$ samples.
The magnetic susceptibility $\chi(T)$ of the
La$_{1-x}$Pr$_x$Pt$_4$Ge$_{12}$ samples becomes temperature-independent
below $\sim$20\,K (see Fig.\,\ref{fig:mag}) and the amplitude simply
scales with the Pr content $x$, as expected for a single-ion CEF effect.
No detectable Curie-like contributions indicating localized magnetic
impurities (e.g.\ Pr$^{3+}$ ions on different crystallographic sites or
in secondary phases) are observed.
The inset of Fig.\ \ref{fig:mag} displays the dependence of $T_c$
on $x$ in La$_{1-x}$Pr$_x$Pt$_4$Ge$_{12}$.
This dependence is weak and the small sagging of the curve below the
linear relationship may be due to the weak crystallographic disorder
introduced by the statistical occupation of the $2a$ site.
Specific heat data (not shown) reveal that also the size of the
specific heat jump $\delta c_p/T_c$ at $T_c$ varies linearly with the Pr
content $x$. This is in contrast to the observations in the series
La$_{1-x}$Pr$_x$Os$_4$Sb$_{12}$ where $\delta c_p/T_c$ shows a strongly
non-linear variation with $x$.\cite{Maple07both} For the substitution series
PrOs$_{4-x}$Ru$_x$Sb$_{12}$ even a strong depression of $T_c$ well below that
of both end members is observed.\cite{Maple07both,LeiShu07,Frederick04}

In the ZF $\mu$SR data (Fig.\ \ref{fig:RlxVsT}) it can be seen that for
PrPt$_4$Ge$_{12}$ the data show the presence of an additional depolarization
below $T_c$. In addition, our data seems to reveal a small dip of $\sigma(T)$
just below $T_c$.
At the moment we do not have an explanation for this dip.
Most plausible would be the
presence of diluted magnetic centers separated on distances of order of the
magnetic penetration depth $\Lambda \simeq 120$\,nm. In such a case,
a reduction of $\sigma$ is expected due to screening of the magnetic
field by the superfluid condensate. The required concentration of
such impurities would be of the order of $\sim 0.01-0.1\%$
[from $\sigma(T>T_c)$].
Clearly, such impurities are not present in our samples as can be
concluded from the absence of an upturn in the magnetic susceptibilities
toward low temperatures (see Fig.\ \ref{fig:mag}).
Another possibility for this dip could be a coupling of Pr nuclei
with free carriers. Below $T_c$ the density of states at the Fermi
level $N_F$ drops. Hence, in case of a Korringa-like coupling it is
expected that the muon relaxation will drop $\propto N_F^2$.
However, the observation of quasi-static magnetism of the Pr nuclei
contradicts this assumption.

Beyond the possible observation of this small dip, the main observation is the
\textit{increase} of $\sigma$ upon lowering the temperature below $T_c$.
Such an increase cannot be explained by a
Pr--Pr RKKY-coupling, since it would reduce the muon
depolarization below $T_c$ ($\propto N_F^n$), in contrast to our
observation. The influence of external fields can be
excluded, since true zero field was controlled with high precision
and, moreover, the Meissner effect automatically shields any fields
in the superconducting state.
Note, only for the heavy-fermion superconductor
PrOs$_4$Sb$_{12}$ a similar spontaneous magnetization was detected
to appear below $T_c$\cite{Aoki03} whereas there is no change of
$\sigma$ in PrRu$_4$Sb$_{12}$ at $T_c$.\cite{Adroja05} In both of
these samples a nearly similar muon depolarization was observed
above $T_c$.
The electronic specific heat coefficient $\gamma$ and $N_F$ are small
for PrPt$_4$Ge$_{12}$.\cite{Gumeniuk08a,Maisuradze09b} In addition,
our measurements of two different samples of PrPt$_4$Ge$_{12}$
(``as-prepared'' and powdered) on two different spectrometers with
the same result and no anomaly at $T_c$ for LaPt$_4$Ge$_{12}$
strongly supports that the enhanced depolarization below $T_c$ is an
intrinsic property of the superconducting state of
PrPt$_4$Ge$_{12}$.

The enhanced muon depolarization below $T_c$ gives evidence of
\textit{time-reversal symmetry} (TRS) breaking in PrPt$_4$Ge$_{12}$. TRS
breaking can be realized for \textit{spin-} or \textit{orbital}
multi-component (vector-) order parameters that may have an internal
phase degree of freedom between the components.\cite{Sigrist91} An
example is the chiral $p$-wave triplet state proposed for
Sr$_2$RuO$_4$\cite{Luke98} and the E$_{2u}$ triplet state for
UPt$_3$.\cite{ThalmeierZwicknagl,Reotier00} Triplet pairing has been
proposed -- and heavily debated -- for
PrOs$_4$Sb$_{12}$.\cite{Higemoto07,Maki04,Chia0305combi} For
PrPt$_4$Ge$_{12}$ we recently reported\cite{Maisuradze09b} that the
superfluid density fits well to the expectations of a chiral $p$-wave
form of the gap function $|\Delta| = \Delta_0|k_x\pm ik_y|$ with a
gap-to-$T_c$ ratio $\Delta_0/T_c = 2.6$ similar to that of
PrOs$_4$Sb$_{12}$.\cite{Maisuradze09b,Maple07both} Most interestingly,
the $T_c$ of PrPt$_4$Ge$_{12}$ is larger than that of other proposed
\textit{spin-triplet} superconductors which have $T_c$ values $<
2.7$\,K.\cite{Heffner90,Luke98,Aoki03,Hillier09}

For LaPt$_4$Ge$_{12}$ we observe no indications for (or an unresolvably
small) TRS breaking. Unfortunately, our investigations of the gap
symmetry are inconclusive at the moment, however a nodeless gap and
spin-singlet pairing has been concluded from NMR relaxation data for
LaPt$_4$Ge$_{12}$.\cite{Toda08a} The weak variation of $T_c$ and of
$\delta c_p/T_c$ with the Pr-content $x$ in
La$_{1-x}$Pr$_x$Pt$_4$Ge$_{12}$ indicates that the order parameters of
the end members are compatible and not separated by a first-order phase
transition. Thus, it is plausible that PrPt$_4$Ge$_{12}$ is also a
\textit{spin-singlet} superconductor. In this case, the observation of
TRS breaking in the condensate requires that the gap function belongs to
a complex \textit{orbitally} degenerate representation leading to an
internal orbital moment of the Cooper pairs. In such a state
supercurrents are induced around nonmagnetic impurities which in turn
generate a condensate magnetic moment density with a spatial extension
of the order of the coherence length $\xi$.\cite{Choi89} Such a complex
spin-singlet state of $T_g$ symmetry with point nodes along the cubic
axes has actually been proposed in Ref.\ \onlinecite{Sergienko04} to
explain the TRS breaking in PrOs$_4$Sb$_{12}$ and as an alternative to
the spin-triplet model. The orbital moment of the Cooper pairs may vary
and in this way the seemingly conflicting observations of a TRS broken
state for PrPt$_4$Ge$_{12}$ and of no visible TRS breaking for the La
compound as well as a continuous changeover in
La$_{1-x}$Pr$_x$Pt$_4$Ge$_{12}$ may appear.

\section{Conclusions}

To conclude, zero field $\mu$SR measurements on PrPt$_4$Ge$_{12}$ and
LaPt$_4$Ge$_{12}$ showed that the dominant contribution for the muon
relaxation comes from the Pr nuclei. Below $T_c$ in PrPt$_4$Ge$_{12}$ we
observe an additional muon depolarization with a temperature variation
resembling that of the superfluid density while no anomalous effect was
seen for LaPt$_4$Ge$_{12}$. This observation indicates TRS breaking in
the superconducting state of PrPt$_4$Ge$_{12}$ with an extraordinary
high $T_c$.\cite{Sigrist91} We have argued that the origin of the TRS
breaking is the unconventional multi-component nature of the order
parameter. From the present experiments no definite conclusion can be
made whether this is due to the spin or orbital degeneracy of the Cooper
pairs. The $T_c$ of 7.8\,K for PrPt$_4$Ge$_{12}$ seems to be rather high
for spin-triplet pairing. In the series La$_{1-x}$Pr$_x$Pt$_4$Ge$_{12}$
the $T_c$ as well as the size of the related specific heat anomaly
vary almost linearly with the Pr content $x$. Together with the absence of
TRS breaking for LaPt$_4$Ge$_{12}$ this renders a spin-triplet Cooper
pairing for these compounds, including PrPr$_4$Ge$_{12}$, unlikely,
since one would expect strong effects for incompatible superconducting
order parameters. Due to the high tetrahedral symmetry, orbital
degeneracies are present which allow for a complex spin-singlet gap
function with an internal phase. Such a kind of pairing with orbital
degeneracy also breaks TRS and may lead to a condensate with a magnetic
moment density.
\newline
Part of this work was performed at the Swiss Muon Source (S$\mu$S), Paul
Scherrer Institute (PSI, Switzerland).


\begin{thebibliography}{99}
\bibitem{RemarkFit} The fit with $\sigma = 0$ and free $\lambda$ is
slightly worse and gives also a temperature-independent relaxation.
\end{thebibliography}
\end{document}